\documentclass{appolb}
\usepackage{amsmath,amssymb}
\usepackage{graphicx}
\usepackage{verbatim}
\usepackage{color}
\usepackage[colorlinks=true]{hyperref}
\usepackage{soul,xcolor}
\usepackage{multirow}

\newcommand{\be}{\begin{equation}}
\newcommand{\ee}{\end{equation}}
\newcommand{\ba}{\begin{eqnarray}}
\newcommand{\ea}{\end{eqnarray}}
\newcommand{\ban}{\begin{eqnarray*}}
\newcommand{\ean}{\end{eqnarray*}}

\begin{document}
\title{Effects of the mean field on fluid dynamics in the relaxation time approximation %
\thanks{Presented by A. Czajka at "Excited QCD 2020", Krynica-Zdr\'oj, Poland, \\ February 2-8, 2020.}%
}

\author{Alina Czajka
\vspace{-0.3cm}
\address{National Centre for Nuclear Research, ul. Pasteura 7, 02-093 Warsaw, Poland}
\\
\vspace{0.3cm}
{Chun Shen}
\vspace{-0.3cm}
\address{Department of Physics and Astronomy, Wayne State University, 
\\
Detroit, MI 48201, USA
\\
Department of Physics, Brookhaven National Laboratory, 
\\
Upton, New York 11973-500, USA}
\\
\vspace{0.3cm}
{Sigtryggur Hauksson, Sangyong Jeon, Charles Gale}
\vspace{-0.3cm}
\address{Department of Physics, McGill University, 3600 rue
University,
\\
Montreal, Quebec H3A 2T8, Canada}
}

\maketitle
\begin{abstract}
In this paper the nonequilibrium correction to the distribution function containing a time and space dependent mass is obtained. Given that, fully consistent fluid dynamic equations are formulated. Then, the physics of the bulk viscosity is elaborated for Boltzmann and Bose-Einstein gases
within the relaxation time approximation. It is found that the parametric form of the ratio $\zeta/\tau_R$ for the quantum gas is affected by the infrared cut-off. This may be an indication that the relaxation time approximation is too crude to obtain a reliable form of bulk viscosity.
\end{abstract}
\PACS{PACS numbers come here}
  
\section{Introduction}

Relativistic viscous hydrodynamics is a very efficient framework to investigate and understand the physics of strongly interacting matter created experimentally in heavy ion collisions \cite{Gale:2013da,Heinz:2013th}. Apart from the conservation laws and constraints on local thermal equilibrium, a viscous hydrodynamical description requires transport coefficients determined by the microscopic structure of a given system. Given that, different phenomena control parametric forms of different coefficients. In weakly interacting systems, the shear viscosity is mostly determined by kinetic energy scale, while bulk viscosity appears as a consequence of the conformal anomaly \cite{Jeon:1994if,Jeon:1995zm,Arnold:2006fz}. Due to the complexity of the symmetry breaking and importance of different energy scales the bulk sector is still much less understood than the shear transport phenomena. In particular, it is important for modelling of heavy ion collisions to have a fluid dynamics formulation where temperature dependent mass is properly included in the bulk sector. This is very challenging in general but doable to some extent in the regime of the coupling constant where analytic methods can be employed. 

In this paper we consider a dilute gas of weakly interacting particles of single species with Bose-Einstein or Boltzmann statistics where effective kinetic theory is applicable and the mean field effects can be systematically examined. Within the kinetic theory many attempts were undertaken so far to provide such a description, see Refs. \cite{Sasaki:2008fg,Chakraborty:2010fr,Bluhm:2010qf,Romatschke:2011qp,Albright:2015fpa,Chakraborty:2016ttq,Tinti:2016bav,Alqahtani:2017jwl} but it seems they were incomplete. Hence we revisited the problem. The entire comprehensive examination of the consequences of the temperature dependent mass on dynamics of the system is presented in our paper~\cite{Czajka:2017wdo}. Here we only provide a very concise summary of the main results. 

\section{Nonequilibrium deviation from the equilibrium distribution
function}
\label{sec-deviation}

The quasiparticle dynamics of a system of a single species is governed by the Boltzmann equation. When the $x$-dependence of the quasiparticle energy is known the equation can be written as follows
\ba
\label{boltz-2}
\big(\tilde k^\mu \partial_\mu -\frac{1}{2} \nabla \tilde m_x^2 \cdot
\nabla_k\big) f=C[f],
\ea
where $C[f]$ is the collision term and $f=f(x,k)$ is a distribution function
of quasiparticles. $\tilde k^\mu=(\tilde k^0, {\bf k})$ is the quasiparticle four-momentum, where $\tilde k_0 \equiv
\mathcal{E}_k$ is the nonequilibrium energy $\mathcal{E}_k = \sqrt{{\bf k}^2+\tilde m_x^2}$. A time and space dependence appears in the mass definition $\tilde m_x^2 \equiv
\tilde m^2(x)=m_0^2+m^2_\text{th}(x)$, where $m_0$ is the constant
mass and $m_\text{th}(x)$ is the nonequilibrium thermal mass, which varies
in time and space. Note that we use tilde and calligraphic letters to denote nonequilibrium quantities. For a system in equilibrium we ommit tilde and use standard letters so that the four-momentum, energy and mass of quasiparticles are denoted by $k^\mu$, $E_k$ and $m_x$, respectively. Also, the thermal mass of quasiparticles in equilibrium is denoted by $m_{\rm eq}$ and the equilibrium phase space density by $f_0$.

The phase-space density function $f(x,k)$ is the main object of the kinetic theory which carries information on the behavior of quasiparticles. When the departure
from the equilibrium state is weak the equilibration process is controlled by the small deviation in the
distribution function
\ba
\label{deltaf-0}
\Delta f(x,k) = f(x,k) - f_0(x,k) =\delta f(x,k) + \delta f_{\text{th}}(x,k),
\ea
where $f_0(x,k)$ is the equilibrium Bose-Einstein distribution function $f_0(x,k) = 1/(\exp\big( E_k(x)\beta(x) \big)-1)$, where $\beta(x)=1/T(x)$ with $T$ being the temperature of the system.
As seen, $\Delta f$ has two parts: $\delta f$ occurs because of hydrodynamic forces, which, in turn, changes the functional form of $f_0$ and $\delta f_{\rm th}=f_{\rm th}-f_0$, which is the effect of a small thermal mass deviation $\Delta m^2_{\rm th}=m^2_{\rm th}-m^2_{\rm eq}$. The function $f_{\rm th}$ has the local-equilibrium form of the Bose-Einstein distribution function $f_{\text{th}}(x,k) \equiv
\left. f_0(x,k)
\right|_{m_0^2 + m^2_{\text{eq}}(x)\to m_0^2 + m^2_{\text{eq}}(x)+\Delta m^2_{\text{th}}(x)}$ and by expanding it one finds the correction $\delta f_{\rm th}$ expressed through $\Delta m_{\rm th}^2$.
Since $\Delta m_{\rm th}^2$ is the nonequilibrium small deviation, which itself is a functional of $\Delta f$, the problem must be solved self-consistently. As a result, one gets
\ba
\label{Delta-f1}
\Delta f = \delta f - T^2 \frac{d m^2_{\text{eq}}}{dT^2}
\frac{f_0(1+f_0)}{E_k}  
\frac{\int dK \delta f}{\int dK E_k f_0(1+f_0)},
\ea
where $dK=d^3{\bf k}/[(2\pi)^3E_k]$. In previous analyses \cite{Sasaki:2008fg,Chakraborty:2010fr,Bluhm:2010qf,Romatschke:2011qp,Albright:2015fpa,Chakraborty:2016ttq,Tinti:2016bav,Alqahtani:2017jwl}, the second term in Eq. (\ref{Delta-f1}) was missing or was incomplete. The thermal mass of the quantum gas is given by $m^2_{\text{eq}} = \lambda T^2/24$, where $\lambda$ is the coupling constant assumed to be samll. The temperature dependence of the thermal mass is found to be $T^2\frac{dm^2_{\text{eq}}}{dT^2} = m^2_{\text{eq}} + T^2 \beta_\lambda/48$, where $\beta_\lambda \equiv  T \frac{d\lambda}{dT}$ is the renormalization group $\beta_\lambda$-function which controls the running of the coupling constant as a function of the energy
scale. $\beta_\lambda$ should be calculated via diagrammatic methods and in case of the scalar theory it is positive and proportional to $\lambda^2$. 

\section{Equations of hydrodynamics with thermal corrections}
\label{sec-boltz-hydro}

The stress-energy tensor of nonequilibrium fluid dynamics takes the
following form
\ba
\label{T-noneq}
T^{\mu\nu} = \int d\mathcal{K} \tilde k^\mu \tilde k^\nu f -g^{\mu\nu}U,
\ea
where $g^{\mu\nu}={\rm diag}(1,-1,-1,-1)$, $U$ is the mean-field contribution and $d\mathcal{K}\equiv d^3{\bf k}/[(2\pi)^3\mathcal{E}_k]$ is the Lorentz invariant measure. First, let us point out that when the system is in equilibrium the stress-energy tensor has the same form as Eq.~(\ref{T-noneq}) but all quantities are replaced by their equilibrium counterparts so that $\tilde k^\mu \to k^\mu$, $\mathcal{E}_k \to E_k$, $f \to f_0$, and $U \to U_0$.  

It is essential to underline that the fluid dynamics equations with thermal effects discussed here are valid as long as all assumptions about quasiparticles of kinetic theory hold. Then, the departure of all quantities from its equilibrium forms is determined by small corrections. In particular one has $f= f_0 + \Delta f$, where $\Delta f$ is given by Eq.~(\ref{Delta-f1}), and $U=U_0 + \Delta U$. The equilibrium mean-field contribution should satisfy $dU_0 = \frac{1}{2} dm^2_{\rm eq} \int dK f_0$ and the nonequilibrium correction is $\Delta U = \frac{1}{2}  \Delta m^2_{\text{th}} \int dk f_0$ to guarantee the energy-momentum conservation law $\partial_\mu T^{\mu\nu}=0$. Consequently, the stress-energy tensor (\ref{T-noneq}) may
be decomposed into the local equilibrium part $T^{\mu\nu}_0$ and the nonequilibrium correction $\Delta T^{\mu\nu}$ as follows
\ba
\label{tensor-delta}
T^{\mu\nu} = T^{\mu\nu}_0 + \Delta T^{\mu\nu}.
\ea
The equilibrium energy-momentum tensor has the familiar form $T_0^{\mu\nu} = \epsilon_0 u^\mu u^\nu - P_0 \Delta^{\mu\nu}$, where $u^\mu$ is the four-velocity and $\Delta^{\mu\nu} = g^{\mu\nu} - u^\mu u^\nu$. $\epsilon_0$ is the energy density and $P_0$ is the local thermodynamic pressure, which are defined as
\ba
\label{energy-pressure}
\epsilon_0 = \int dK \, E_k^2 f_0 - U_0, \qquad\qquad 
P_0 = \frac{1}{3} \int dK \, {\bf k}^2 f_0 + U_0.
\ea
The presence of the mean-field contribution in Eq.~(\ref{energy-pressure}) does not change the enthalpy, $h_0=\epsilon_0 + P_0$. One can also check that the thermodynamic relation $T s_0= TdP_0/dT=\epsilon_0 + P_0$, where $s_0$ is the entropy density, is fulfilled. 

$\Delta T^{\mu\nu}$, which depends on $\Delta f$ and $\Delta U$, carries entire dynamical
information needed to determine how the nonequilibrium system
evolves into its equilibrium state. The Landau matching is defined by the eigenvalue problem, which in the
fluid rest frame can be expressed by the
conditions on the energy and the momentum densities $T^{00} = \epsilon$ and $T^{0i} = 0$, respectively.
Given that, one defines the local equilibrium as the state having the same local energy and the momentum density, which is the essence of the Landau matching conditions found as
\ba
\label{T-00-landau}
\Delta T^{00} =\int dK \bigg[E_k^2 - T^2 \frac{dm^2_{\text{eq}}}{dT^2} \bigg] \delta f = 0,
\qquad
\label{T-0i-landau}
\Delta T^{0i} =\int dK E_k k^i  \Delta f=  0.
\ea
$\Delta T^{ij}$ can be manipulated and reorganized in such a way to separate
the spin 0 part and the spin 2 part, $\Delta T^{ij} = \pi^{ij} + \delta^{ij} \Pi$, where the shear-stress tensor $\pi^{ij}$ and the bulk pressure $\Pi$ have commonly known forms
\ba 
\label{pi-ij}
\pi^{ij} = \int dK \Big( k^i k^j - \frac{1}{3}\delta^{ij}{\bf k}^2 \Big)
\delta f, \qquad\qquad
\Pi = \frac{1}{3} \int dK {\bf k}^2 \delta f.
\ea

\section{Transport coefficients in the Anderson-Witting model}
\label{sec-bulk-ce}

In the Anderson-Witting model the Boltzmann equation with the $x$-dependent thermal mass is given by
\be
\left(k^\mu\partial_\mu 
- {1\over 2}\partial_i m_{\rm eq}^2 
{\partial\over \partial k_i}\right)f_0(x,k)
= -{E_k\over \tau_R} \Delta f(x,k),
\ee
where $k^\mu = (E_k, {\bf k})$ and $\tau_R$ is the relaxation time which is assumed to be energy independent. $\Delta f$ is the nonequilibrium correction given by Eq.~(\ref{Delta-f1}) and we let $\delta f = f_0(1+f_0)\phi$, where $\phi = \phi_{\rm s} + \phi_{\rm b}$, that is, it consist of the shear and bulk part. Solving the Anderson-Witting model, one finds their forms to be
\ba
\label{eq:phi_s}
\phi_{\rm s}(k)
&=& 
-{\tau_R \over TE_k} \Big( k^i k^j -\frac{1}{3} \delta^{ij} {\bf k}^2 \Big) \partial_j  u_i ,
\\
\label{eq:phi_b}
\phi_{\rm b}(k) &=&
\tau_R \beta(\partial_i u^i)(c_s^2 - 1/3)
 \left(E_k -{1\over E_k}{J_{3,0} - T^2(dm_{\rm eq}^2/dT^2)J_{1,0}\over J_{1,0} 
- T^2(dm_{\rm eq}^2/dT^2) J_{-1,0}}
\right).\qquad
\ea
where $c_s^2$ is the speed of sound and the factor $c_s^2-1/3$ depending both on the mass $m_0^2$ and $\beta_\lambda$ fixes the nonconformality parameter. The thermodynamic functions $J_{n,q}$ are defined as follows $J_{n,q} = 1/(2q+1)!! \int dK  (u \cdot k)^{n-2q} (-\Delta_{\mu\nu} k^\mu k^\nu )^q \, f_0(k)(1+f_0(k))$. One can check that with these forms of solution the energy of the system is conserved and the Landau matching conditions are satisfied. Having given the solutions (\ref{eq:phi_s}) and (\ref{eq:phi_b}) one can use Eq.~(\ref{pi-ij}) to find shear-stress tensor and bulk pressure. Next by comparing them with $\pi^{ij} = 2\eta \sigma^{ij}$, where $\sigma^{ij}=-1/2(\partial^i u^j + \partial^j u^i -2/3 g^{ij} \partial_k u^k)$ and $\Pi = -\zeta \partial_i u^i$ the ratios $\eta/\tau_R$ and $\zeta/\tau_R$ can be extracted. Therefore, from the shear part one finds the known form of the ratio $\eta/\tau_R = (\epsilon_0+P_0)/5$
and from the bulk part one gets
\ba
\label{eq:zeta_be}
{\zeta \over\tau_R}
&\approx& 
T^4 \left({1\over 3} - c_s^2\right)^2 
\left( {2\pi^3 T \over 25 m_{x}} - {4\pi^2 \over 75 } 
\left( 1-\frac{9 m^2_{\rm eq}}{8 m^2_x} \right) \right),
\ea
where $m_x=\sqrt{m_0^2 + m_{\rm eq}^2(x)}$. The ratio for the Boltzmann statistics $f_{0,c}(k) = e^{-\beta E_k}$ can be found analogously and it is
\be
\label{zeta-boltzmann}
{\zeta_{\rm Boltz}\over \tau_R}
\approx
T^4 \left({1\over 3} - c_s^2\right)^2
\left(
{60\over \pi^2} - {36m_{x} \over \pi T}
\right).
\ee

Note that the structure of the expression (\ref{eq:zeta_be}) is slightly different than the one in (\ref{zeta-boltzmann}) because fo the factor $T/m_{x}$. The origin of this difference comes from
the fact that the infrared limit of the Bose-Einstein factor behaves like
$f_0(k)\sim T/E_k$ while the Boltzmann factor does not show such a behavior. 

\section{Conclusions}
\label{sec-summary}

In this paper we examined the effects of mean field on fluid dynamics. We found the correction to the distribution function which enabled us to formulate fully consistent equations of fluid dynamics as well as to solve the Anderson-Witting model to compute $\zeta/\tau_R$ of Bose-Einstein and Boltzmann gases. The ratio $\zeta/\tau_R$ for the the Boltzmann gas has a parametrically expected form, that is, it is given by the nonconformality parameter squared. In case of the Bose-Einstein gas, the leading order term of $\zeta/\tau_R$ has an additional energy scale dependent factor $T/m_x$. We suspect that it is an indication that the relaxation time approximation applied here is too crude to get the expected form of the ratio since the constant relaxation time is insensitive to the soft scale.

\section*{Acknowledgments}

This work is supported in part by the National Science Centre, Poland, under grant 2018/29/B/ST2/00646 and by the Natural Sciences and Engineering Research Council of Canada.

\end{document}